\documentclass[twocolumn,secnumarabic,amssymb, nobibnotes, aps, prl,superscriptaddress]{revtex4-1}

\usepackage{graphicx}
\usepackage{dcolumn}
\usepackage{bm}
\usepackage{hyperref}
\usepackage{natbib}
\usepackage{amsmath}
\usepackage{amsfonts}

\begin{document}
	
	\preprint{APS/123-QED????????????}
	
	\title{Plasmon Mediated Near Field Heat Transfer in Double Layer Graphene}
	\title{Plasmonic Tuning of the Near Field Heat Transfer in Double Layer Graphene}
	
	\author{Xuzhe Ying}
	\affiliation{School of Physics and Astronomy, University of Minnesota, Minneapolis, MN 55455, USA}
	
	\author{Alex Kamenev}
	\affiliation{School of Physics and Astronomy, University of Minnesota, Minneapolis, MN 55455, USA}
		\affiliation{William I. Fine Theoretical Physics Institute,  University of Minnesota,
Minneapolis, MN 55455, USA}


	
	
	\begin{abstract}
	We discuss the non-radiative heat transfer in non-equilibrium double layer graphene system. We show that at the neutrality point the heat exchange is dominated by the inter-layer plasmon modes and derive analytic expressions for the heat current  as a function of temperature and the interlayer separation. These results show that for a range of low temperatures the two graphene layers are much more efficient heat exchanger than conventional metals. The physical reason behind this phenomenon is
the presence of inter-band excitations with a large energy, and a small momentum in graphene spectrum. Plasmonic mechanism of the heat transfer is sharply 
suppressed by electrostatic doping. This allows for tuning of the heat exchange by a small applied voltage between the two layers.  	\end{abstract}
	
	\pacs{1111}
	\maketitle

	Heat transfer between two bodies may greatly exceed the black-body radiation limit in the near field regime. The effect is due to the non-radiative,  evanescent modes of the electromagnetic field \cite{pendry1999radiative}, which can transfer energy between closely spaced dielectric  \cite{PhysRevB.4.3303,PhysRevB.50.18517,pendry1999radiative,RevModPhys.79.1291}, or metallic  \cite{PhysRevB.95.115427,levin1980theory,kamenev2018near,wise2020role} surfaces. Specifically in metals, 
the near field heat transfer (NFHT) is associated with the Coulomb interactions between thermally excited electron-hole pairs  \cite{PhysRevB.97.195450,PhysRevB.96.155437,PhysRevB.95.115427}. The resonant modes of the Coulomb coupled electron-hole plasmas -- the surface plasmons -- may be expected to play an important role in this process, offering a higher efficiency  and tunability of the heat transfer  \cite{mulet2001nanoscale,volokitin2003adsorbate,PhysRevB.69.045417}. Yet in conventional metals, plasmons play a modest role, dominating the heat transport   
only in a  narrow range of temperatures and scattering rates  \cite{wise2020role}.

%

Graphene is a promising material \cite{novoselov2004electric,novoselov2005two}, which allows for a high degree of control over its plasmon spectra, through electrostatic doping, temperature and choice of the spacer \cite{grigorenko2012graphene,koppens2011graphene,low2014graphene}. Due to the Dirac point in its electronic spectrum, graphene supports unusual plasmon modes \cite{PhysRevB.75.205418,PhysRevB.80.205405,PhysRevLett.97.266406,PhysRevB.83.155441}, which could serve as mediators of the inter-layer heat transfer.  The role of plasmons in NFHT between two graphene layers was highlighted in Refs.~[\onlinecite{PhysRevB.85.155418}] and [\onlinecite{PhysRevB.85.155422}], who presented numerical evidence for the effect. It was shown that NFHT between the two graphene layers can be greatly enhanced by a factor of $10^2\sim10^3$ compared to the black body radiation at room temperature. 

	
	In this article, we take a kinetic approach \cite{kamenev2011field,PhysRevB.96.155437,PhysRevB.97.195450,wise2020role}. The energy transfer is captured by the energy exchange during the electron-electron scattering and the processes of the creation and annihilation of the electron-hole pairs in the two layers. This approach allows us to analyze the problem analytically and extract the parametric dependence (such as temperature, $T$, and interlayer distance, $d$) for various heat transfer channels. Here we focus on the plasmon channel, which dominates NFHT in graphene through a broad range of the relevant parameters. 
	
		\begin{figure}[tb]
		\centering
		\includegraphics[width=\linewidth]{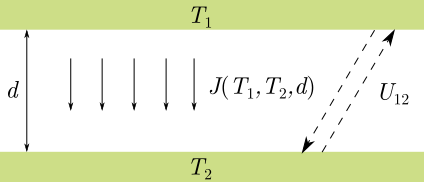}
		\caption{Schematic system setup: two layers of graphene are at different temperatures $T_1$ and $T_2$ and separated by a distance $d$. The interlayer electron interaction is $U_{12}$. The heat flux is $J(T_1,T_2,d)$.}
		\label{Fig:SysSetup}
	\end{figure}

\begin{figure}[tb]
		\includegraphics[scale=0.5]{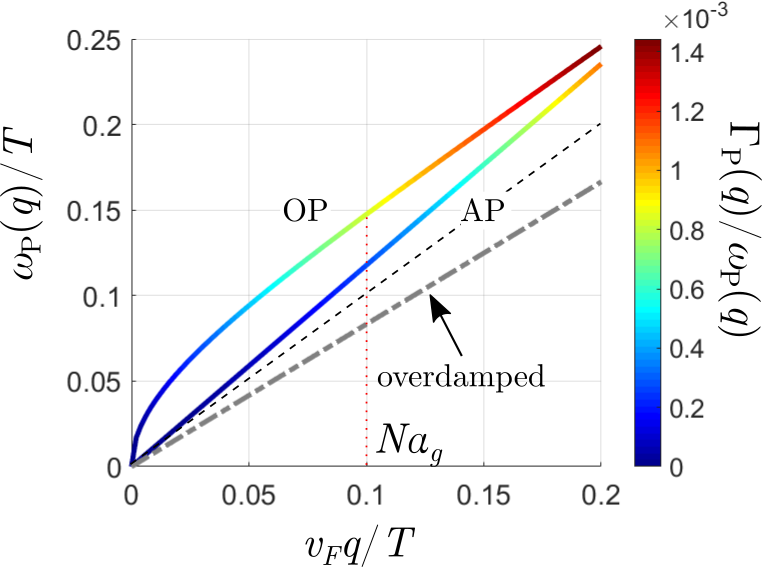}
		\caption{Frequencies, Eq.~(\ref{Eq:PlasmonFrequecy}),  and decay rates, Eq.~(\ref{Eq:PlasmonDecayZfactor}), of the OP and AP modes. Here the two graphene layers are taken at the same temperature, $T=20v_F/d$ and $N\alpha_g = 0.1$. 
The grey dot-dashed line shows the overdamped AP at $T=10v_F/d$, for comparison. Black dashed line is $\omega =v_Fq$.		
}
		\label{Fig:Plasmon}
	\end{figure}

\begin{figure}[tb]
		\centering
		\includegraphics[width=\linewidth]{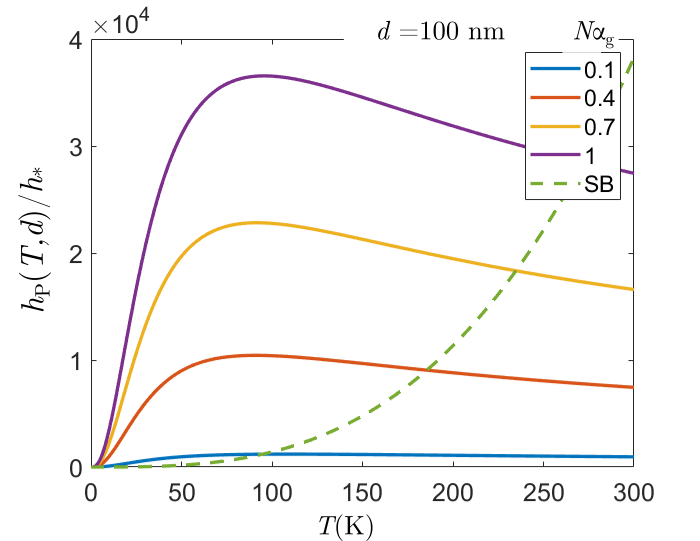}
		\caption{Plasmon contribution to the near field heat transfer conductance at various values of fine structure constant $N\alpha_g$ at a distance $d=100$ nm. There are three temperature regimes at a fixed layer separation, Eq.~(\ref{Eq:HeatCond}): low temperature, intermediate temperature and high temperature. The green dashed line corresponds to the black body radiation, given by Stefan-Boltzmann (SB) law.}
		\label{Fig:HeatCond}
	\end{figure}

	We consider two graphene layers kept at different temperatures as shown in Fig.~\ref{Fig:SysSetup}. The electrons in the two layers interact through the screened Coulomb potential. As a result, the electron density fluctuations of the two graphene layers are strongly coupled. This leads to the two plasmon branches - the optical and acoustic plasmons, corresponding to the in-phase and the out-of-phase oscillations of the electron densities respectively. Their dispersion relations  are shown in Fig.~\ref{Fig:Plasmon}. The plasmon mediated NFHT conductance as a function of temperature is summarized in Fig.~\ref{Fig:HeatCond}.  At small temperature it exhibits $T^3$ dependence, which is similar to the radiative Stefan--Boltzmann law. The coefficient however is greatly exceeding the Stefan--Boltzmann one, as the the speed of light (in the negative two power) is substituted with the graphene Fermi velocity, $v_F$.  The thermal conductance riches a broad maximum at $T\approx v_F/d$ and slowly decrease as $T^{-1/2}$ at higher temperatures.

Such non-monotonic thermal conductance is a signature of an undoped graphene. In a doped case the maximum is sharply suppressed  once the 
chemical potential, $\mu$, exceeds the characteristic temperature $v_F/d$. This opens a way to manipulate the heat transport by purely electrostatic 
means. Indeed, applying a small voltage between the two graphene layers, or between a graphene layer and a gate, one can induce a charge density and thus 
non-zero chemical potential.    	We show that a minute voltage is sufficient to decrease the heat transfer exponentially.

\begin{figure}[tb]
		\centering
		\includegraphics[width=\linewidth]{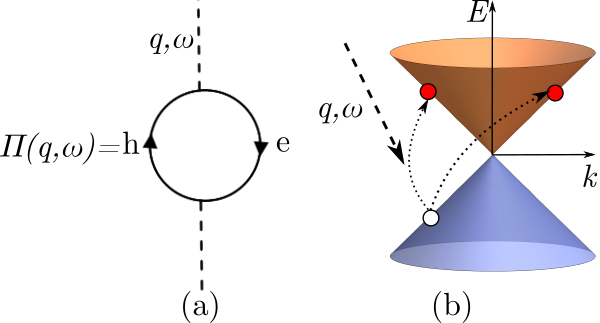}
		\caption{(a) Feynman diagram for the polarization operator $\Pi(q,\omega)$; (b) Inter-band electron-hole excitations excited by a virtual (evanescent) photon with a high frequency $\omega>v_Fq$.}
		\label{Fig:PiAndSPE}
	\end{figure}

In the second order in the dynamically screened interlayer Coulomb interactions (RPA approximation), the net heat flux between the two layers with temperatures $T_1$ and $T_2$  is given by \cite{pendry1999radiative,PhysRevB.95.115427,doi:10.1146/annurev-conmatphys-031016-025203,PhysRevB.96.155437,PhysRevB.97.195450,kamenev2018near,caroli1971direct}:
	\begin{equation}
	\begin{split}
	J(T_1,T_2,d)=\int &\frac{d\omega d^2q}{(2\pi)^3}[N_1(\omega)-N_2(\omega)]\\
	\times&\Im\Pi_1(q,\omega)\Im\Pi_2(q,\omega)|U_{12}(q,\omega)|^2, 
	\end{split}
	\label{Eq:HeatFlux}
	\end{equation}
where the integration runs over the 2D momentum, $\boldsymbol{q}$, and energy, $\omega$. Here $N_j(\omega)=\omega/\left[\exp(\omega/T_j)-1\right]$ and $\Pi_j(q,\omega)$ are the Planck function and the polarization operator, Fig.~\ref{Fig:PiAndSPE}(a), with $j=1,2$ labeling the layers. Finally, $U_{12}(q,\omega)$ is the screened interlayer Coulomb interaction, which may be written as:
	\begin{equation}
	U_{12}(q,\omega)=\frac{1}{\epsilon_{12}(q,\omega)}\, \frac{2\pi v_F\alpha_g}{q}\, e^{-qd},
	\end{equation}
where $v_F$ and $\alpha_g=e^2/(\hbar v_F\tilde \epsilon)$ are the Fermi velocity of the Dirac spectrum and the fine structure constant in graphene, respectively;  $\tilde \epsilon$ is the dielectric constant of a spacer between the two layers and $d$ is the interlayer separation. Within the RPA approximation the interlayer dielectric function, $\epsilon_{12}(q,\omega)$, is given by the solution of the $2\times 2$ matrix  Dyson's equation  \cite{altland2010condensed,PhysRevB.23.805}:
	\begin{eqnarray}
						\label{Eq:ScreeningFunction}
	\epsilon_{12}(q,\omega)=&&\, 1+\frac{2\pi v_F\alpha_g}{q}\left[\Pi_1(q,\omega)+\Pi_2(q,\omega)\right]\\
	+&&\, \left(\frac{2\pi v_F\alpha_g}{q}\right)^2\Pi_1(q,\omega)\Pi_2(q,\omega)(1-e^{-2qd}).   \nonumber
	\end{eqnarray}
	
	As explained below, the main contribution to NFHT in graphene is coming from the resonances (poles) of the dynamically screened interactions, $U_{12}(q,\omega)$  \cite{mulet2001nanoscale,volokitin2003adsorbate,PhysRevB.69.045417,wise2020role}. These plasmon resonances at, in general, complex frequency, $ \omega=\omega_{\text{P}}(q)-i\Gamma_{\text{P}}(q)$, 
are determined by  zeros of the dielectric function, $	\epsilon_{12}(q,\omega)=0$. In a vicinity of the plasmon pole the screened interaction takes the form 
\begin{equation}
						\label{Eq:U12-pole}
U_{12}(q,\omega)=\frac{Z(q)}{\omega-\omega_{\text{P}}(q)+i\Gamma_{\text{P}}(q)}\, (2\pi v_F\alpha_g)e^{-qd}, 
\end{equation}
where $\omega_{\text{P}}(q)$ and $\Gamma_{\text{P}}(q)$  are the plasmon frequency and the decay rate, and $Z(q)$ is the residue of the plasmon propagator.
For the underdamped plasmons, with $\Gamma_{\text{P}}(q)\ll \omega_{\text{P}}(q)$, one can evaluate the frequency integral in Eq.~(\ref{Eq:HeatFlux}) in the pole approximation \cite{PhysRevB.52.14796,RevModPhys.88.025003} to single out the plasmon contribution to the heat flux:
	\begin{eqnarray}
							\label{Eq:PlasmonHeatFlux}
	&&J_{\text{P}}(T_1,T_2,d)=\!\int\!\! \frac{d^2q}{(2\pi)^2}[N_1(\omega_{\text{P}}(q))-N_2(\omega_{\text{P}}(q))]\\
	&&\times\, \Im\Pi_1(q,\omega_{\text{P}}(q))\Im\Pi_2(q,\omega_{\text{P}}(q))\, \frac{Z^2(q)(2\pi v_F\alpha_g)^2 e^{-2qd}}{2\,\Gamma_{\text{P}}(q)}. \nonumber
	\end{eqnarray}
	
It is clear from here that the plasmon part of the heat flux is weighted by $ \Im\Pi_j(q,\omega_{\text{P}}(q))	$. This is the reason why plasmons 
do not play a role in conventional metals at low temperatures  \cite{kamenev2018near,wise2020role,PhysRevB.95.115427,PhysRevLett.18.546}. Indeed, the 2D plasmon frequency,  $\omega_\text{P}(q)\propto \sqrt{q}$ 
 \cite{giuliani2005quantum} satisfies 	$\omega_\text{P}(q)\gg v_Fq$. On the other hand, at $T=0$ all particle-hole excitations have $\omega\leq v_Fq$, resulting in 
 $ \Im\Pi_j(q,\omega_{\text{P}}(q))=0$. At a finite temperature  $ \Im\Pi_j(q,\omega_{\text{P}}(q))\propto \exp\{-\epsilon_F/T\}$  \cite{PhysRevLett.18.546}  and thus the plasmon contribution is still small. It may become significant, however at $T> v_F/l$ and $d\sim l$, in disordered metals with $l$ being the mean free path  \cite{wise2020role}. 
 
Undoped graphene is qualitatively different.  Due to the inter-band transitions, Fig.~\ref{Fig:PiAndSPE}(b), there are particle-hole excitations with $\omega \geq v_Fq$, i.e. 
$ \Im\Pi_j(q,\omega_\text{P}(q))\neq 0$, even at $T=0$. This is the reason why the plasmon contribution, Eq.~(\ref{Eq:PlasmonHeatFlux}), dominates the heat transfer in the undoped graphene bilayers. To make the quantitative predictions let us review the 
plasmon dispersion and damping in graphene bilayers.

We are interested in the frequency range $v_F q<\omega\lesssim T_j$. The first inequality specifies the range where underdamped plasmons may exist. The second one follows from the fact that the Planck functions in Eq.~(\ref{Eq:PlasmonHeatFlux}) limit the frequency integration by the temperature. In this range, the graphene polarization operator, Fig.~\ref{Fig:PiAndSPE}(a), is given by \cite{PhysRevB.83.155441}:
	\begin{equation}
						\label{Eq:polarization}
	\begin{split}
	\Pi_j(q,\omega)=&N\frac{T_j}{v_F^2}\frac{\ln 2}{\pi}\left[1-\frac{\omega}{\sqrt{\omega^2-(v_Fq)^2}}\right]\\
	&+i\frac{N}{16v_F^2}\frac{(v_Fq)^2\tanh(\omega/4T_j)}{\sqrt{\omega^2-(v_Fq)^2}},
	\end{split}
	\end{equation}
where $N$ is the Dirac cone degeneracy ($N=4$ for graphene due to two spins and two valleys). As mentioned above, the high frequency, $\omega>v_Fq$,  polarization operator has a finite imaginary part. This is a direct result of having the inter-band electron-hole excitations near the Dirac point, shown in Fig.~\ref{Fig:PiAndSPE}(b). Hence, graphene is a good emitter/absorber of high frequency modes, such as plasmons.

The double layer system supports two modes called \cite{PhysRevB.80.205405}  optical (OP) and acoustic plasmon (AP).
Their complex frequencies follow from $\epsilon_{12}(q,\omega)=0$, with the dielectric function given by Eq.~(\ref{Eq:ScreeningFunction}) and the polarization operators given by Eq.~(\ref{Eq:polarization}). As detailed in the Supplemental Material  \cite{SupplementalM}, their dispersions are, Fig.~\ref{Fig:Plasmon}:
	\begin{subequations}
	\begin{align}
	&\omega_{\text{OP}}(q)=\frac{\left[v_Fq+2\ln 2\ N\alpha_g(T_1+T_2)\right]\sqrt{v_Fq}}{\sqrt{v_Fq+4\ln 2\ N\alpha_g(T_1+T_2)}}  \label{Eq:OPFrequency};\\
	&\omega_{\text{AP}}(q)=v_Fq\sqrt{2\ln 2\ N\alpha_g\frac{T_1T_2}{T_1+T_2}\frac{d}{v_F}}\label{Eq:APFrequency}. 
	\end{align}
	\label{Eq:PlasmonFrequecy}
	\end{subequations}
These expressions hold for $q<d^{-1}$, while in the opposite limit Eqs.~(\ref{Eq:PlasmonFrequecy}) smoothly connect to the plasmon frequencies of each individual graphene layer  \cite{PhysRevLett.97.266406,PhysRevB.83.155441}.

	The OP corresponds to in-phase oscillations of the electron densities. Its dispersion relation, Eq.~(\ref{Eq:OPFrequency}), is similar to the plasmon frequency of a single graphene layer \cite{PhysRevLett.97.266406,PhysRevB.83.155441}.  For   $v_Fq<N\alpha_g(T_1+T_2)$, the OP exhibits a square root dispersion, $\omega_{\text{OP}}(q)\approx \sqrt{\ln 2 N\alpha_g(T_1+T_2)v_Fq}$; while at larger momenta, $v_Fq>N\alpha_g(T_1+T_2)$, it has a linear spectrum $\omega_{\text{OP}}(q)\approx  v_Fq$, Fig.~\ref{Fig:Plasmon}.
The AP mode corresponds to out-of-phase, charge neutral oscillations, thus the linear dispersion,  Eq.~(\ref{Eq:APFrequency}), Fig.~\ref{Fig:Plasmon}. Notice that the AP is only under-damped at a high enough temperature, so that its velocity is larger than $v_F$ \cite{PhysRevB.80.205405}, $T_{1,2}>v_F(N\alpha_g d)^{-1}$. In the opposite limit the AP mode merges with the intra-band single particle excitations and is over-damped.
	
	The plasmon decay rate, $\Gamma_{\text{P}}(q)$, and the residue of the plasmon propagator, $Z(q)$, can be written as:
	\begin{subequations}
		\begin{align}
		&\Gamma_{\text{P}}(q)=\frac{\pi}{64 \ln2}\left[\omega_{\text{P}}^2(q)-(v_Fq)^2\right]\frac{\omega_{\text{P}}(q)}{T_1T_2}; \\
		&Z(q)=\frac{s_{\text{P}}}{2\ln2\ N\alpha_g}\frac{v_F}{T_1+T_2}\frac{\left[\omega_{\text{P}}^2(q)-(v_Fq)^2\right]^{3/2}}{(v_Fq)^2},
		\end{align}
		\label{Eq:PlasmonDecayZfactor}
	\end{subequations}
where  $s_{\text{P}}=s_{\text{OP}/AP}=\pm 1$ is a sign related to the plasmon species. The ratio between the plasmon decay rate and its frequency is much less than one, Fig.~(\ref{Fig:Plasmon}), through the entire range, $\omega_{\text{P}}(q)\lesssim T$, contributing to the heat current.  Therefore, the plasmon modes are indeed underdamped, justifying   Eq.~(\ref{Eq:PlasmonHeatFlux}).
	



	Employing Eq.~(\ref{Eq:PlasmonHeatFlux}) along with Eqs.~(\ref{Eq:PlasmonFrequecy})-(\ref{Eq:PlasmonDecayZfactor}), one  finds the plasmon contribution to NFHT. We present results, Fig.~\ref{Fig:HeatCond}, for the heat transfer conductance \cite{ilic2018active}:
	\begin{equation}
	h_{\text{P}}(T,d)=\lim_{T_{1,2}\rightarrow T}\frac{J_{\text{P}}(T_1,T_2,d)}{T_1-T_2}, 
	\label{Eq:HeatCondDef}
	\end{equation}
which can be measured through the temperature relaxation time: $\tau(T,d)=C(T)/2h_{\text{P}}(T,d)$, where $C(T)$ is the specific heat of the undoped graphene  \cite{pop2012}. In the limit $N\alpha_g\ll 1$, one can identify the three temperature regimes, where the asymptotic analytic expressions are available in Supplemental Material \cite{SupplementalM}, cf. Fig.~\ref{Fig:HeatCond}, 
	\begin{equation}
	h_{\text{P}}(T,d)\!=\!  \left\{ \hskip -.15cm \begin{array}{l}
	\text{\small $ 1.49\!\times\!10^{-1}$} \frac{(N\alpha_g)^2}{v_F^2}\, T^3;  \ \ \ \ \ T<\frac{v_F}{d}, \\ \\
	\text{\small $ 2.71\!\times\!10^{-3} $} \frac{(N\alpha_g)^2}{v_F^2} \frac{v_F^3}{d^3};  \ \ \ \ \  \frac{v_F}{d}<T<\frac{v_F}{N\alpha_gd}, \\ \\
	\text{\small $1.29\!\times\!10^{-3}$} \frac{(N\alpha_g)^2}{v_F^2}\frac{(v_F/d)^{7/2}}{(N\alpha_g T)^{1/2}};  \ \ \ \ T>\frac{v_F}{N\alpha_gd}. 
	\end{array} \right.
	\label{Eq:HeatCond}
	\end{equation}
At low temperatures, $T< v_F/(N\alpha_gd)$, only OP mode is underdamped, while in the opposite limit both OP and AP modes are underdamped and yield comparable contributions  to NFHT conductance. For $T< v_F/d$ the integral in Eq.~(\ref{Eq:PlasmonHeatFlux}) is cut by the Planck function,  limiting the relevant momenta by  $v_Fq<T$. As a result, the heat transfer conductance follows $h\sim (N\alpha_g)^2T^3/v^2_F$ -- the Stefan-Boltzmann like relation (the latter describes the far-field radiation and is given by $h_\mathrm{SB}\sim T^3/c^2$). There is no dependence on the separation, $d$, between the layers in this regime. In the intermediate temperature range, the inverse layer separation, $d^{-1}$, effectively limits the momentum integral, $v_Fq<v_Fd^{-1}<T$ making the heat conductance temperature independent, but  $\propto (v_F/d)^{3}$ instead.  Notice that to observe the high temperature regime of Eq.~(\ref{Eq:HeatCond}), it's necessary to have  $N\alpha_g>(v_F/c)^{2/5}\approx 0.1$, otherwise the Stefan-Boltzmann far-field radiation mechanism is more important than NFHT, Fig.~\ref{Fig:HeatCond}.

We have found that NFHT conductance in the undoped graphene double layers follows a characteristic non-monotonic temperature dependence, Fig.~\ref{Fig:HeatCond}. It reaches a broad maximum at $T_m\approx v_F/d$ and slowly decreases at higher temperature, before the radiative transfer 
kicks in at $\approx 4T_m(N\alpha_g)^{3/7}$. 
Throughout the low temperature range, $T\leq T_m$, the NFHT in graphene greatly exceeds that in conventional metals \cite{kamenev2018near,wise2020role,PhysRevB.95.115427}. The corresponding enhancement factor is $(N\alpha_g \kappa d)^2\gg 1$, where $\kappa\gtrsim 1nm^{-1}$ is the inverse screening radius in metals. In fact, at the maximum, $T\approx T_m$, the graphene NFHT is only a factor of $(N\alpha_g)^2$ off the Pendry's boundary \cite{pendry1999radiative} for the maximal possible 
heat flux.  

The plasmon contribution to the heat flux can be switched off by electrostatic gating. Notice that a voltage, $V$, applied between the two graphene layers induces the charge densities $n=\pm \tilde \epsilon\, V/(ed)$ in the two layers, measured from the charge neutrality point. The chemical potential of each graphene layer changes to $\pm\mu$, accordingly. There is a characteristic gate voltage, $V_{\text{T}}=0.23 \times \frac{N\alpha_g}{e}\frac{dT^2}{v_F}$, when the chemical potential is equal to the temperature, $\mu=T$. For a smaller gate voltage, $V\lesssim V_{\text{T}}$, the chemical potential is smaller than temperature, $\mu\approx T\times V/V_{\text{T}} < T$. In this case, the plasmon contribution to the NFHT, Eq.~(\ref{Eq:HeatCond}) is basically intact. At a larger gate voltage, $V\gtrsim V_{\text{T}}$, the chemical potential is greater than the temperature, $\mu\approx1.7\times T\sqrt{V/V_{\text{T}}}> T$. In this case, the imaginary part of the polarization operator, $\Im\Pi(q,\omega_P(q))\sim\exp\left[-\mu/T\right]$, is exponentially small \cite{PhysRevB.75.205418}. As discussed before, the heat transfer conductance is proportional to the imaginary part of the polarization operator, $h\sim\left(\Im \Pi\right)^2/\Gamma_\text{P}$,  (where $\Gamma_\text{P}\sim \Im \Pi$). Hence, the plasmon contribution to the NFHT is exponentially small, $h\sim\exp\left[-1.7\sqrt{V/V_{\text{T}}}\right]$ for $V> V_{\text{T}}$. Therefore the plasmon contribution to NFHT is effectively  switched off by the gate voltage $V_{\text{T}}$. For a typical inter-layer separation $d=100\text{ nm}$, it is given by $V_{\text{T}}=0.026\times\left(\frac{T}{10\text{K}}\right)^2\text{ meV}$. Thus the small electrostatic potential may be used to tune on and off the heat flux in the graphene double layers.

	We are grateful to D. Basko, I. Gorny, A. Levchenko, O. Ilic, M. Sammon, J. Schmalian, and H. Yeh for useful discussions. 
	This research was supported by NSF Grant No. DMR-1608238 and partly by the Heising-Simons Foundation, the Simons Foundation, and NSF Grant No. NSF PHY-1748958. 
	
	\bibliography{PlasmonNFHTRef}{}

\end{document}


\preprint{APS/123-QED????????????}
	
	\title{Supplemental Material: Plasmonic Tuning of the Near Field Heat Transfer in Double Layer Graphene}
	
	\author{Xuzhe Ying}
	\affiliation{School of Physics and Astronomy, University of Minnesota, Minneapolis, MN 55455, USA}
	
	\author{Alex Kamenev}
	\affiliation{School of Physics and Astronomy, University of Minnesota, Minneapolis, MN 55455, USA}
	\affiliation{William I. Fine Theoretical Physics Institute,  University of Minnesota,
Minneapolis, MN 55455, USA}

	\begin{abstract}
	\end{abstract}
	

	

	\pacs{1111}
	\maketitle

	\section{Basic Formalism}
	
	In this section, we derive the general equation for the plasmon contribution to the heat flux between two graphne layers.
	
	Start with the general equation for near field heat transfer \cite{kamenev2018near,wise2020role}: 
	\begin{equation}
	\begin{split}
	J=\int\frac{d\omega d^2q}{(2\pi)^3}&[N_1(\omega)-N_2(\omega)]
	\\ \times&
	\Im\Pi_1(q,\omega)\Im\Pi_2(q,\omega)|U_{12}(q,\omega)|^2, 
	\end{split}
	\end{equation}
	where $\Pi_j(q,\omega)$ is the polarization operator and $N_j(\omega)=\omega/\left[\exp(\omega/Tji)-1\right]$ is the Planck function with temperature $T_j$, with $j=1,2$ labelling the graphene layers. The screened interlayer Coulomb interaction $U_{12}(q,\omega)$ has the following generic form:
	\begin{equation}
	U_{12}(q,\omega)=\frac{2\pi v_F\alpha_g}{q\epsilon_{12}(q,\omega)}e^{-qd}
	\end{equation}
	where $q$ and $\omega$ is the transferred momentum (in 2D) and frequency. $\epsilon_{12}(q,\omega)$ is the dynamical screening function. As will be discussed in the next section, the interaction is screened by electrons in both layers of graphene. As a result, the plasmon modes will be strongly coupled. 
	
	Around plasmon frequency the screened interlayer interaction could be rewritten as:
	\begin{equation}
	U_{12}(q,\omega)=\frac{2\pi v_F\alpha_g Z(q)}{\omega-\omega_{\text{P}}(q)+i\Gamma_{\text{P}}(q)}e^{-qd}
	\end{equation}
	The plasmon frequency $\omega_{\text{P}}(q)$ is determined by:
	\begin{equation}
	\Re \left[q\epsilon_{12}(q,\omega_{\text{P}}(q))\right]=0
	\label{Eq:DefPlasmon}
	\end{equation}
	Meanwhile, the decay rate of plasmon modes is given by:
	\begin{equation}
	\Gamma_{\text{P}}(q)=Z(q)\Im\left[q\epsilon_{12}(q,\omega_{\text{P}}(q))\right]
	\label{Eq:DecayRate}
	\end{equation}
	with the $Z$ factor being the inverse of the derivative of the real part of $\Re\left[q\epsilon_{12}(q,\omega)\right]$:
	\begin{equation}
	Z(q)=\left[\frac{\partial\Re\left[q\epsilon_{12}(q,\omega)\right]}{\partial\omega}\right]^{-1}\lvert_{\omega=\omega_{\text{P}}(q)}
	\end{equation}

	The interaction $U_{12}(q,\omega)$ is presumably highly peaked at the plasmon frequency. The absolute value squared of interaction $|U_{12}(q,\omega)|^2$ can be considered as a delta function times a regular function, $|U_{12}(q,\omega)|^2\propto\delta(\omega-\omega_{\text{P}}(q))$ \cite{RevModPhys.88.025003,PhysRevB.52.14796}. As a result, the plasmon contribution to the heat flux can be singled out to be:
	\begin{equation}
	\begin{split}
	&J_{\text{P}}=\int\frac{d^2q}{2(2\pi)^2}[N_1(\omega_{\text{P}}(q))-N_2(\omega_{\text{P}}(q))]\\
	\times&\Im\Pi_1(q,\omega_{\text{P}}(q))\Im\Pi_2(q,\omega_{\text{P}}(q))\frac{(2\pi v_F\alpha_g)^2Z^2(q)e^{-2qd}}{\Gamma_{\text{P}}(q)}
	\end{split}
	\end{equation}
	
	To be more specific, consider the denominator of the interaction $|U_{12}(q,\omega)|^2$ around plasmon frequency:
	\begin{equation}
	\begin{split}
	\frac{1}{\left[\omega-\omega_{\text{P}}(q)\right]^2+\Gamma_{\text{P}}^2(q)}
	=&\frac{\left[\omega-\omega_{\text{P}}(q)\right]^2+\Gamma_{\text{P}}^2(q)}{\left(\left[\omega-\omega_{\text{P}}(q)\right]^2+\Gamma_{\text{P}}^2(q)\right)^2}\\
	\approx&\frac{\Gamma_{\text{P}}^2(q)}{\left(\left[\omega-\omega_{\text{P}}(q)\right]^2+\Gamma_{\text{P}}^2(q)\right)^2}
	\end{split}
	\end{equation}
	In the second line, the term $\left[\omega-\omega_{\text{P}}(q)\right]^2$ in the numerator is neglected around plasmon frequency. Notice the following relation in the limit $\Gamma_{\text{P}}(q)\rightarrow 0$:
	\begin{equation}
	\frac{\Gamma_{\text{P}}(q)}{\left[\omega-\omega_{\text{P}}(q)\right]^2+\Gamma_{\text{P}}^2(q)}\rightarrow \pi \delta(\omega-\omega_{\text{P}}(q))
	\end{equation}
	Therefore, the denominator of $|U_{12}(q,\omega)|^2$ is given by:
	\begin{equation}
	\begin{split}
	\frac{1}{\left[\omega-\omega_{\text{P}}(q)\right]^2+\Gamma_{\text{P}}^2(q)}
	\approx&\frac{\Gamma_{\text{P}}(q)\pi\delta(\omega-\omega_{\text{P}}(q))}{\left[\omega-\omega_{\text{P}}(q)\right]^2+\Gamma_{\text{P}}^2(q)}\\
	=&\ \frac{\pi\delta(\omega-\omega_{\text{P}}(q))}{\Gamma_{\text{P}}(q)}
	\end{split}
	\end{equation}
	The interaction $|U_{12}(q,\omega)|^2$ is thus given by:
	\begin{equation}
	|U_{12}(q,\omega)|^2\approx(2\pi v_F\alpha_g)^2 Z^2(q)e^{-2qd}\frac{\pi\delta(\omega-\omega_{\text{P}}(q))}{\Gamma_{\text{P}}(q)}
	\end{equation}
	Substituting this expression into the equation for near filed heat transfer and performing the frequency integration, we were able to separate out the plasmon contribution to the heat transfer.

	\section{Plasmon Modes}

	In this section, we discuss plasmon modes of double layer graphene \cite{PhysRevB.75.205418,PhysRevB.80.205405}. The collective motion of electrons in the two layers of graphene are strongly coupled by the Coulomb interaction. Two plasmon modes are expected: one is the usual charge plasmon, called optical plasmon; the other one is called acoustic plasmon, related to the out of phase fluctuation of the charge densities. 
	
	As discussed in the main text, the screening function is determined with random phase approximation (RPA)\cite{altland2010condensed,PhysRevB.23.805}:
	\begin{widetext}
		\begin{equation}
		\epsilon_{12}(q,\omega)=1+\frac{2\pi v_F\alpha_g}{q}\left[\Pi_1(q,\omega)+\Pi_2(q,\omega)\right]
		+\left(\frac{2\pi v_F\alpha_g}{q}\right)^2\Pi_1(q,\omega)\Pi_2(q,\omega)\left(1-e^{-2qd}\right)
		\end{equation}
	\end{widetext}
	In addition, we consider small momentum $v_Fq\ll 2T_{1,2}$ and large frequency $v_Fq<\omega<T_{1,2}$, where the plasmon modes are under damped. The upper limit $\omega<T_{1,2}$ is set implicitly by the Planck function in the equation of near field heat transfer. In this region, the polarization operator is given by\cite{PhysRevB.83.155441}:
	\begin{equation}
	\begin{split}
	\Pi_j(q,\omega)=&N\frac{T_j}{v_F^2}\frac{\ln 2}{\pi}\left[1-\frac{\omega}{\sqrt{\omega^2-(v_Fq)^2}}\right]\\
	&+i\frac{N}{16v_F^2}\frac{(v_Fq)^2\tanh(\omega/4T_j)}{\sqrt{\omega^2-(v_Fq)^2}}
	\end{split}
	\end{equation}
	where $N=4$ accounts two spins and two valleys.

	To obtain the plasmon frequency in the long wavelength limit, we assumed the imaginary part of the polarization operator to be close to zero $\Im\Pi_i(q,\omega)\approx 0^+$. This is a reasonable approximation, in the sense that the imaginary part of the polarization operator is indeed smaller than the real part by a factor of:
	\begin{equation}
	\frac{\Im\Pi_i(q,\omega)}{\Re\Pi_i(q,\omega)}\sim\frac{v_Fq}{T_i}\frac{(v_Fq)\tanh(\omega/4T_i)}{\omega-\sqrt{\omega^2-(v_Fq)^2}}
	\end{equation}
	When frequency is large $\omega\gg v_Fq$, this ratio is approximately $(\omega/T_i)^2\ll 1$; when the frequency is small $\omega\sim v_Fq$, this ratio is approximately $(v_Fq)^3/(T_i^2\omega)\ll 1$. Therefore, the imaginary part of the polarization operator can be neglected temporarily in determining the plasmon frequency. It is important in determining the plasmon decay rate.
	
	The plasmon frequency was found by solving $\Re(q\epsilon_{12}(q,\omega))=0$:
	\begin{widetext}
		\begin{equation}
		0=q+N(2\pi v_F\alpha_g)\frac{T_1+T_2}{v_F^2}\frac{\ln 2}{\pi}\left[1-\frac{\omega}{\sqrt{\omega^2-(v_Fq)^2}}\right]\\
		+N^2(2\pi v_F\alpha_g)^2\frac{T_1T_2}{v_F^4}\left(\frac{\ln 2}{\pi}\right)^2\left[1-\frac{\omega}{\sqrt{\omega^2-(v_Fq)^2}}\right]^2\frac{1-e^{-2qd}}{q}
		\end{equation}

		Two solutions can be found generically:
		\begin{equation}
		1-\frac{\omega_{\pm}}{\sqrt{\omega_{\pm}^2-(v_Fq)^2}}=\frac{-(T_1+T_2)\pm\sqrt{(T_1+T_2)^2-4T_1T_2(1-e^{-2qd})}}{2N(2\pi \alpha_g)\frac{T_1T_2}{v_Fq}\frac{\ln 2}{\pi}(1-e^{-2qd})}
		\end{equation}

		Simple analytical expression for the two branches of the plasmon frequencies can be found when we considered the long wavelength limit $q<d^{-1}$ (namely, $1-e^{-2qd}\approx 2qd$). In the long wavelength limit, the above equation simplifies to:
		\begin{equation}
		\begin{split}
		&1-\frac{\omega_{+}}{\sqrt{\omega_{+}^2-(v_Fq)^2}}=-\frac{v_Fq}{2\ln 2\ N\alpha_g(T_1+T_2)}\\
		&1-\frac{\omega_{-}}{\sqrt{\omega_{-}^2-(v_Fq)^2}}=-\frac{(T_1+T_2)}{4\ln 2\ N\alpha_gT_1T_2}\frac{v_F}{d}
		\end{split}
		\end{equation}
		The explicit solution for the frequencies are presented below as two branches of plasmon modes: optical plasmon (OP) and acoustic plasmon (AP) \cite{PhysRevB.23.805,PhysRevB.80.205405}.
		
		The optical plasmon corresponds to the usual charge density fluctuation. The electron density in two layers of graphene exhibits in-phase oscillation. The optical plasmon frequency can be obtained by taking the $\omega_+$ solution:
		\begin{equation}
		\omega_{\text{OP}}(q)=\frac{\left[v_Fq+2\ln 2\ N\alpha_g(T_1+T_2)\right]\sqrt{v_Fq}}{\sqrt{v_Fq+4\ln 2\ N\alpha_g(T_1+T_2)}}
		\end{equation}
		Meanwhile, the acoustic plasmon mode corresponds to the fluctuation in the relative density. Its frequency is given by taking the $\omega_-$ solution:
		\begin{equation}
		\omega_{\text{AP}}(q)=v_Fq\sqrt{2\ln 2\ N\alpha_g\frac{T_1T_2}{T_1+T_2}\frac{d}{v_F}}
		\end{equation}
		
		Now, we turn to derive the expressions for the $Z(q)$ factors and the decay rates $\Gamma_{\text{P}}(q)$ for both plasmon branches. The $Z(q)$ factor is given by the derivative of $\Re(q\epsilon_{12}(q,\omega))$:
		
		\begin{equation}
		\begin{split}
		Z^{-1}(q)=&\left.\frac{\partial \Re(q\epsilon_{12}(q,\omega))}{\partial\omega}\right|_{\omega=\omega_{\text{P}}(q)}\\
		=&N(2\pi v_F\alpha_g)\frac{T_1+T_2}{v_F^2}\frac{\ln 2}{\pi}\frac{(v_Fq)^2}{(\omega^2-(v_Fq)^2)^{3/2}}\\
		+&\left.N^2(2\pi v_F\alpha_g)^2\frac{T_1T_2}{v_F^4}\left(\frac{\ln 2}{\pi}\right)^22\left[1-\frac{\omega}{\sqrt{\omega^2-(v_Fq)^2}}\right]\frac{(v_Fq)^2}{(\omega^2-(v_Fq)^2)^{3/2}}\frac{1-e^{-2qd}}{q}\right|_{\omega=\omega_{\text{P}}(q)}
		\end{split}
		\end{equation}

		Substituting the solution of plasmon frequencies, one can find the $Z(q)$ factor. For optical plasmon, the $Z(q)$ factor is:
		\begin{equation}
		\begin{split}
		Z^{-1}_{\text{OP}}(q)=2\ln 2\ N\alpha_g\frac{T_1+T_2}{v_F}\frac{(v_Fq)^2}{\left[\omega^2_{\text{OP}}(q)-(v_Fq)^2\right]^{3/2}}
		=2\ln 2\ N\alpha_g\frac{T_1+T_2}{v_F}\frac{(v_Fq)^2}{\omega^3_{\text{OP}}(q)}\left(1+\frac{v_Fq}{2\ln 2\ N\alpha_g(T_1+T_2)}\right)^3
		\end{split}
		\end{equation}
		Meanwhile, for acoustic plasmon:
		\begin{equation}
		\begin{split}
		Z^{-1}_{\text{AP}}(q)=-2\ln 2\ N\alpha_g\frac{T_1+T_2}{v_F}\frac{(v_Fq)^2}{\left[\omega^2_{\text{AP}}(q)-(v_Fq)^2\right]^{3/2}}
		=-2\ln 2\ N\alpha_g\frac{T_1+T_2}{v_F}\frac{(v_Fq)^2}{\omega^3_{\text{AP}}(q)}\left(1+\frac{T_1+T_2}{4\ln 2\ N\alpha_gT_1T_2}\frac{v_F}{d}\right)^3
		\end{split}
		\end{equation}
		As discussed in the main text, the acoustic plasmon is only under-damped in the high temperature regime $T_{1,2}>v_F(N\alpha_gd)^{-1}$. As a result, the $Z^{-1}_{\text{AP}}(q)$ can be further simplified to be:
		\begin{equation}
		Z^{-1}_{\text{AP}}(q)=-2\ln 2\ N\alpha_g\frac{T_1+T_2}{v_F}\frac{(v_Fq)^2}{\omega^3_{\text{AP}}(q)}
		\end{equation}

		To end this section, we show the calculation of the plasmon decay rates. At this step, one should not neglect the imaginary part of the polarization operator. The decay rate is defined as $\Gamma_{\text{P}}(q)=Z(q)\Im\left[q\epsilon_{12}(q,\omega_{\text{P}}(q))\right]$, where the imaginary part of $q\epsilon_{12}(q,\omega)$ is given by:
		\begin{equation}
		\begin{split}
		\Im\left[q\epsilon_{12}(q,\omega)\right]=&\frac{2\pi N\alpha_g}{16v_F}\frac{(v_Fq)^2}{\sqrt{\omega^2-(v_Fq)^2}}\left[\tanh(\omega/4T_1)+\tanh(\omega/4T_2)\right]\\
		&+q\frac{(2\pi N\alpha_g)^2}{16}\frac{\ln 2}{\pi}\left[1-\frac{\omega}{\sqrt{\omega^2-(v_Fq)^2}}\right]\frac{\left(1-e^{-2qd}\right)}{\sqrt{\omega^2-(v_Fq)^2}}\left[T_1\tanh(\omega/4T_2)+T_2\tanh(\omega/4T_1)\right]
		\end{split}
		\end{equation}
		As before, one needs to substitute in the plasmon frequency. In addition, it's convenient to consider low frequency situation $\omega<T_{1,2}$ due to the implicit constraint from the Planck function, so that $\tanh(\omega/4T_{1,2})\approx\omega/4T_{1,2}$.  Also long wavelength limit is assumed such that $1-e^{-2qd}\approx 2qd$. For optical plasmon, one can find:
		\begin{equation}
		\begin{split}
		\Im\left[q\epsilon_{12}(q,\omega_{\text{OP}}(q))\right]=\frac{2\pi N \alpha_g}{64v_F}\frac{(v_Fq)^2\omega_{\text{OP}}(q)}{\sqrt{\omega^2_{\text{OP}}(q)-(v_Fq)^2}}\left[\frac{1}{T_1}+\frac{1}{T_2}\right]
		=\frac{2\pi N\alpha_g}{64v_F}(v_Fq)^2\left(\frac{v_Fq}{2\ln 2\ N\alpha_g(T_1+T_2)}+1\right)\left[\frac{1}{T_1}+\frac{1}{T_2}\right]
		\end{split}
		\end{equation}
		For acoustic plasmon:
		\begin{equation}
		\begin{split}
		\Im\left[q\epsilon_{12}(q,\omega_{\text{AP}}(q))\right]=-\frac{2\pi N\alpha_g}{64v_F}\frac{(v_Fq)^2\omega_{\text{AP}}(q)}{\sqrt{\omega^2_{\text{AP}}(q)-(v_Fq)^2}}\frac{T_1+T_2}{T_1T_2}\left[\frac{T_1}{T_2}+\frac{T_2}{T_1}-1\right]
		\approx-\frac{2\pi N\alpha_g}{64v_F}(v_Fq)^2\frac{T_1+T_2}{T_1T_2}\left[\frac{T_1}{T_2}+\frac{T_2}{T_1}-1\right]
		\end{split}
		\end{equation}
		
		Combining the result for $Z(q)$ factor and $\Im\left[q\epsilon(q,\omega)\right]$, it's straightforward to find the decay rates of plasmon modes. For optical plasmon:
		\begin{equation}
		\begin{split}
		\Gamma_{\text{OP}}(q)=\frac{\pi}{64\ln 2}\left[\omega^2_{\text{OP}}(q)-(v_Fq)^2\right]\frac{\omega_{\text{OP}}(q)}{T_1T_2}
		=\frac{\pi}{64\ln 2T_1T_2}\frac{\omega^3_{\text{OP}}(q)}{\left(\frac{v_Fq}{2\ln 2\ N\alpha_g(T_1+T_2)}+1\right)^2}
		\end{split}
		\end{equation}
		For acoustic plasmon:
		\begin{equation}
		\begin{split}
		\Gamma_{\text{AP}}(q)=\frac{\pi}{64\ln 2}\left[\omega^2_{\text{AP}}(q)-(v_Fq)^2\right]\frac{\omega_{\text{AP}}(q)}{T_1T_2}\left[\frac{T_1}{T_2}+\frac{T_2}{T_1}-1\right]
		=\frac{\pi}{64\ln 2}\frac{\omega^3_{\text{AP}}(q)}{T_1T_2}\left[\frac{T_1}{T_2}+\frac{T_2}{T_1}-1\right]
		\end{split}
		\end{equation}

		All the results in this section are summarized in the TABLE~\ref{Tab:PlasmonProperties}, where temperatures are taken to be equal $T_1=T_2=T$.\\
		\newline
		\begin{table}
			\begin{tabular}{ccc}
				
				\    & \ \ \ \  Optical Plasmon (OP)\ \ \ \  &\ \ \ \  Acoustic Plasmon (AP)\ \ \ \ \\
				\hline
				\ & \ & \ \\
				Frequency $\omega_{\text{P}}(q)$ & $\frac{\left[v_Fq+4\ln 2\ N\alpha_gT\right]\sqrt{v_Fq}}{\sqrt{v_Fq+8\ln 2\ N\alpha_gT}}$ & $v_Fq\sqrt{\ln 2\ N\alpha_gT\frac{d}{v_F}}$\\
				\  & \ &\ \\
				\ \ \ \ \ \ $Z^{-1}(q)$ Factor\ \ \ \ \ \  & \ \ \ \ \ \ $4\ln 2\ N\alpha_g\frac{T}{v_F}\frac{(v_Fq)^2}{\left[\omega^2_{\text{OP}}(q)-(v_Fq)^2\right]^{3/2}}$ \ \ \ \ \ \ \ & $-4\ln 2\ N\alpha_g\frac{T}{v_F}\frac{(v_Fq)^2}{\left[\omega^2_{\text{AP}}(q)-(v_Fq)^2\right]^{3/2}}$\\
				\ &\ &\ \\
				Decay Rate $\Gamma_{\text{P}}(q)$ & $\frac{\pi}{64\ln 2}\left[\omega^2_{\text{OP}}(q)-(v_Fq)^2\right]\frac{\omega_{\text{OP}}(q)}{T^2}$ & $\frac{\pi}{64\ln 2}\left[\omega^2_{\text{AP}}(q)-(v_Fq)^2\right]\frac{\omega_{\text{AP}}(q)}{T^2}$\\
				\ &\ &\ \\
				\hline
			\end{tabular}
			\caption{Summary of the properties of the plasmon modes: frequency, $Z$ factor and decay rate. Here, the temperatures of two graphene layers are taken to be equal to $T$.}
			\label{Tab:PlasmonProperties}
		\end{table}
		
	\end{widetext}

	\section{Heat Conductance}
	
	In this section, we show the calculation of the near field heat transfer conductance, which is defined as:
	\begin{equation}
	h_{\text{P}}(T,d)=\lim_{T_{1,2}\rightarrow T}\frac{J_{\text{P}}(T_1,T_2,d)}{T_1-T_2}
	\end{equation}
	The explicit equation would be:
	\begin{equation}
	\begin{split}
	h_{\text{P}}(T,d)=\int&\frac{d^2q}{2(2\pi)^2}\frac{dN(\omega_{\text{P}}(q))}{dT}
	\left[\Im\Pi(q,\omega_{\text{P}}(q))\right]^2\\
	&\times\frac{(2\pi v_F\alpha_g)^2Z^2(q)e^{-2qd}}{\Gamma_{\text{P}}(q)}
	\end{split}
	\end{equation}
	The result can be divided into three regimes: the low temperature regime ($\text{I}$), the intermediate temperature regime ($\text{II}$) and the high temperature regime ($\text{III}$). We discuss them separately in each subsection below.

	\subsection{High Temperature Regime}
	
	In this subsection, we consider temperature to be high:
	\begin{equation}
	T>\frac{v_F}{N\alpha_gd}>\frac{v_F}{d}
	\end{equation}
	At high temperature, both optical and acoustic plasmon exist. 
	
	In particular, in the long wavelength limit $q<d^{-1}<N\alpha_gT/v_F$, the optical plasmon frequency has a square root dependence on the momentum:
	\begin{equation}
	\omega_{\text{OP}}(q)=\sqrt{2\ln 2\ N \alpha_gTv_Fq}
	\label{Eq:OPFrequencySQRT}
	\end{equation}
	Notice that $\omega_{\text{OP}}(q)\gg v_Fq$. Therefore, the decay rate can be approximated as:
	\begin{equation}
	\Gamma_{\text{OP}}(q)=\frac{\pi}{64\ln 2}\frac{\omega^3_{\text{OP}}(q)}{T^2}
	\label{Eq:OPDecaySqrt}
	\end{equation}
	At the same time, the Z factor is given by:
	\begin{equation}
	Z_{\text{OP}}(q)=\frac{1}{4\ln 2\ N\alpha_g}\frac{v_F}{T}\frac{\omega^3_{\text{OP}}(q)}{(v_Fq)^2}
	\label{Eq:ZFactorSqrt}
	\end{equation}
	The imaginary part of the polarization operator is:
	\begin{equation}
	\begin{split}
	\Im\Pi_i(q,\omega_{\text{OP}}(q))=&\frac{N}{16v_F^2}\frac{(v_Fq)^2}{\omega_{\text{OP}}(q)}\tanh(\omega_{\text{OP}}(q)/4T)\\
	\approx&N\frac{(v_Fq)^2}{64v_F^2T}
	\end{split}
	\end{equation}
	In the second line, $\omega_{\text{OP}}(q)\ll T$ is assumed. With this assumption, one can make a simplification on the Planck function with $\omega_{\text{OP}}(q)\ll T$:
	\begin{equation}
	\frac{dN(\omega)}{dT}=\frac{\omega^2}{4T^2}\frac{1}{\sinh^2\frac{\omega}{2T}}\approx 1
	\end{equation}
	
	Putting the equations above, one can find the optical plasmon contribution to the heat conductance:
	\begin{equation}
	\begin{split}
	h_{\text{OP}}(T,d)=&\int\frac{d^2q}{2(2\pi)^2}\frac{\pi}{2^8\cdot\ln 2}\frac{\omega^3_{\text{OP}}(q)}{T^2}e^{-2qd}\\
	=&\int\frac{d^2q}{2(2\pi)^2}\frac{\pi}{2^8\cdot\ln 2}\frac{(2\ln 2\ N\alpha_gTv_Fq)^{3/2}}{T^2}e^{-2qd}\\
	=&\frac{15}{2^{15}}\sqrt{\pi\ln 2}(v_FN\alpha_g)^{3/2}\frac{1}{d^{7/2}}\frac{1}{\sqrt{T}}
	\end{split}
	\end{equation}
	
	One could do the same process for acoustic plasmon. The acoustic plasmon frequency is:
	\begin{equation}
	\omega_{\text{AP}}(q)=v_Fq\sqrt{\ln 2\ N\alpha_gT\frac{d}{v_F}}
	\end{equation}
	The $Z$ factor is given by:
	\begin{equation}
	Z_{\text{AP}}(q)=-\frac{1}{4\ln 2\ N\alpha_g}\frac{\omega^3_{\text{AP}}(q)}{(v_Fq)^2}\frac{v_F}{T}
	\end{equation}
	The decay rate is given by:
	\begin{equation}
	\Gamma_{\text{AP}}(q)=\frac{\pi}{64\ln2}\frac{\omega^3_{\text{AP}}(q)}{T^2}
	\end{equation}
	We assumed the same for the imaginary part of polarization operator, $\Im\Pi_j(q,\omega_{\text{AP}}(q))\approx\frac{(v_Fq)^2}{64v_F^2T}$, and the Planck function, $\frac{dN(\omega)}{dT}\approx 1$. Putting the equations above together, one could fine the acoustic plasmon contribution to the heat conductance:
	\begin{equation}
	\begin{split}
	h_{\text{AP}}(T,d)=&\int\frac{d^2q}{2(2\pi)^2}\frac{\pi}{2^8\cdot\ln 2}\frac{\omega^3_{\text{AP}}(q)}{T^2}e^{-2qd}\\
	=&\int\frac{d^2q}{2(2\pi)^2}\frac{\pi(v_Fq)^3}{2^8\cdot\ln 2}\frac{(\ln 2\ N\alpha_gTd/v_F)^{3/2}}{T^2}e^{-2qd}\\
	=&\frac{3}{2^{12}}\sqrt{\ln 2}(v_FN\alpha_g)^{3/2}\frac{1}{d^{7/2}}\frac{1}{\sqrt{T}}
	\end{split}
	\end{equation}
	
	Summing the optical and acoustic plasmon contribution to he heat conductance, one finds the near field heat transfer conductance in the high temperature regime:
	\begin{equation}
	\begin{split}
	h_{\text{III}}(T,d)=&\frac{15\sqrt{\pi}+24}{2^{15}}\sqrt{\ln 2}(v_FN\alpha_g)^{3/2}\frac{1}{d^{7/2}}\frac{1}{\sqrt{T}}\\
	\sim& (v_FN\alpha_g)^{3/2}\frac{1}{d^{7/2}}\frac{1}{\sqrt{T}}
	\end{split}
	\end{equation}

	\subsection{Intermediate Regime}
	
	In this subsection, we consider intermediate temperature regime:
	\begin{equation}
	\frac{v_F}{d}<T<\frac{v_F}{N\alpha_gd}
	\end{equation}
	At this temperature regime, only optical plasmon exists.
	
	The momentum integration in the near field heat transfer conductance was done in two steps for small momentum, $0<q<4\ln 2\ N\alpha_gT/v_F$, and large momentum $q>4\ln 2\ N\alpha_gT/v_F$. 
	
	For small momentum region $0<q<4\ln 2\ N\alpha_gT/v_F$, the optical plasmon properties fully parallel the discussion in the high temperature regime. As a result, the heat conductance is given by the following equation:
	\begin{equation}
	h_{\text{OP},<}(T,d)=\int_{0}^{q^*}\frac{d^2q}{2(2\pi)^2}\frac{\pi}{2^8\cdot\ln 2}\frac{\omega^3_{\text{OP}}(q)}{T^2}e^{-2qd}
	\end{equation}
	where the upper limit of integration is $q^*=4\ln 2\ N\alpha_gT/v_F<d^{-1}$. Hence the exponential factor can be neglected:
	\begin{equation}
	\begin{split}
	h_{\text{OP},<}(T,d)=&\int_{0}^{q^*}\frac{d^2q}{2(2\pi)^2}\frac{\pi}{2^8\cdot\ln 2}\frac{\omega^3_{\text{OP}}(q)}{T^2}\\
	=&\frac{(\ln 2)^3}{7\sqrt{2}}(N\alpha_g)^5\frac{T^3}{v_F^2}
	\end{split}
	\end{equation}
	
	At large momentum region $q>4\ln 2\ N\alpha_gT/v_F$, the optical plasmon frequency is approximately:
	\begin{equation}
	\omega_{\text{OP}}(q)=v_Fq\left[1+\frac{1}{2}\left(\frac{4\ln 2\ N\alpha_gT}{v_Fq}\right)^2\right]
	\label{Eq:OP_Frequency_Largeq}
	\end{equation}
	It's straightforward to see the following relation:
	\begin{equation}
	\left(\omega_{\text{OP}}(q)\right)^2-(v_Fq)^2=\left(4\ln 2\ N\alpha_gT\right)^2
	\label{Eq:OP_Frequency_Largeq1}
	\end{equation}
	With the above relation, we found the $Z$ factor is given by:
	\begin{equation}
	Z_{\text{OP}}(q)=v_F\frac{\left(4\ln 2\ N\alpha_gT\right)^2}{(v_Fq)^2}
	\label{Eq:OP_Z_Largeq}
	\end{equation}
	The decay rate is given by:
	\begin{equation}
	\Gamma_{\text{OP}}(q)=\frac{\pi\ln 2}{4}\left(N\alpha_g\right)^2v_Fq
	\label{Eq:OP_Gamma_Largeq}
	\end{equation}
	The imaginary part of the polarization operator is taken to be:
	\begin{equation}
	\begin{split}
	\Im\Pi_i(q,\omega_{\text{OP}}(q))=&\frac{N}{16v_F^2}\frac{(v_Fq)^2\tanh(\omega_{\text{OP}}(q)/4T)}{\sqrt{\omega_{\text{OP}}^2(q)-(v_Fq)^2}}\\
	=&\frac{N}{64v_F^2}\frac{(v_Fq)^3}{4\ln 2 \ N\alpha_gT^2}
	\end{split}
	\label{Eq:OP_ImPi_Largeq}
	\end{equation} 
	The derivative of Planck function is assumed to be $\frac{dN(\omega)}{dT}\approx 1$. Putting together the equations above, we found the heat transfer conductance to be:
	\begin{equation}
	\begin{split}
	h_{\text{OP},>}(T,d)=&\int_{q^*}\frac{d^2q}{2(2\pi)^2}\frac{\pi\ln 2}{2^4}(N\alpha_g)^2v_Fqe^{-2qd}\\
	=&\frac{\ln 2}{2^9}(N\alpha_g)^2\frac{v_F}{d^3}(2+4q^*d)e^{-2q^*d}
	\end{split}
	\end{equation}
	The lower integration limit is $q^*=4\ln 2\ N\alpha_gT/v_F\ll d^{-1}$. As a result, the heat conductance in the large momentum region is approximately:
	\begin{equation}
	h_{\text{OP},>}(T,d)\approx\frac{\ln 2}{2^8}(N\alpha_g)^2\frac{v_F}{d^3}
	\end{equation}
	
	It turns out that the major contribution comes from the large momentum region:
	\begin{equation}
	\frac{h_{\text{OP},>}(T,d)}{h_{\text{OP},<}(T,d)}\sim\left(\frac{v_F/d}{N\alpha_gT}\right)^3>1
	\end{equation}
	Therefore, in the intermediate temperature regime, the near field heat transfer conductance is given by:
	\begin{equation}
	\begin{split}
	h_{\text{II}}(T,d)=&\frac{\ln 2}{2^8}(N\alpha_g)^2\frac{v_F}{d^3}\\
	\sim&(N\alpha_g)^2\frac{v_F}{d^3}
	\end{split}
	\end{equation}
	
	\subsection{Low Temerature Regime}
	
	In this subsection, we consider low temperature regime:
	\begin{equation}
	T<\frac{v_F}{d}
	\end{equation}
	At this temperature regime, only optical plasmon exists. The analysis completely parallels the previous subsection of intermediate temperature regime. The only difference is that the momentum of interest is $0<q<T/v_F$, where the upper limit is given by the Planck function.
	
	At small momentum $q<4\ln 2\ N\alpha_gT/v_F$, the heat conductance is given by:
	\begin{equation}
	h_{\text{OP},<}(T,d)=\frac{(\ln 2)^3}{7\sqrt{2}}(N\alpha_g)^5\frac{T^3}{v_F^2}
	\end{equation}
	
	At larger momentum $4\ln 2\ N\alpha_gT/v_F<q<T/v_F(<d^{-1})$, the momentum integration in the equation of heat conductance is cut by the derivative of Planck function:
	\begin{equation}
	\begin{split}
	h_{\text{OP},>}(T,d)=\int_{q^*}&\frac{d^2q}{2(2\pi)^2}\frac{dN(\omega)}{dT}
	N^2\left[\Im\Pi(q,\omega)\right]^2\\
	&\times\frac{(2\pi v_F\alpha_g)^2Z^2(q)}{\Gamma_{\text{P}}(q)}\rvert_{\omega=\omega_{\text{P}}(q)}
	\end{split}
	\end{equation}
	where the exponential factor $e^{-2qd}$ from the interaction is neglected. The derivative of Planck function is:
	\begin{equation}
	\frac{dN(\omega)}{dT}=\frac{\omega^2}{4T^2}\frac{1}{\sinh^2\frac{\omega}{2T}}
	\end{equation}
	Substitute in the expressions for the $Z$ factor and plasmon decay rate in Eq.~(\ref{Eq:OP_Frequency_Largeq})-(\ref{Eq:OP_Gamma_Largeq}), while the imaginary part of the polarization operator should keep the $\tanh$ term:
	\begin{equation}
	\begin{split}
	\Im\Pi_i(q,\omega_{\text{OP}}(q))=&\frac{N}{16v_F^2}\frac{(v_Fq)^2\tanh(\omega_{\text{OP}}(q)/4T)}{\sqrt{\omega_{\text{OP}}^2(q)-(v_Fq)^2}}\\
	=&\frac{N}{16v_F^2}\frac{(v_Fq)}{4\ln 2 \ N\alpha_gT^2}\tanh\left(\frac{v_Fq}{4T}\right)
	\end{split}
	\end{equation}, one finds the heat conductance to be:
	\begin{equation}
	\begin{split}
	h_{\text{OP},>}(T,d)=&\int_{q^*}\frac{d^2q}{2(2\pi)^2}\pi\ln 2(N\alpha_g)^2\frac{v_Fq}{4}\frac{\tanh^2\frac{v_Fq}{4T}}{\sinh^2\frac{v_Fq}{2T}}\\
	=&\frac{(\pi^2 - 6)\ln 2 }{18}(N\alpha_g)^2\frac{T^3}{v_F^2}\left(1-\mathcal{O}(N\alpha_g)^3\right)
	\end{split}
	\end{equation}
	where the lower integration limit is $q^*=4\ln 2\ N\alpha_gT$. The fine structure constant is assumed to be a small factor. Hence, the heat conductance is given by:
	\begin{equation}
	h_{\text{OP},>}(T,d)=\frac{(\pi^2 - 6)\ln 2 }{18}(N\alpha_g)^2\frac{T^3}{v_F^2}
	\end{equation}
	
	Still, the major contribution comes from the large momentum region:
	\begin{equation}
	\frac{h_{\text{OP},>}(T,d)}{h_{\text{OP},<}(T,d)}\sim\frac{1}{(N\alpha_g)^3}\gg 1
	\end{equation}
	Therefore, in the low temperature regime, the near field heat transfer conductance is given by:
	\begin{equation}
	\begin{split}
	h_{\text{I}}(T,d)=&\frac{(\pi^2 - 6)\ln 2 }{18}(N\alpha_g)^2\frac{T^3}{v_F^2}\\
	\sim&(N\alpha_g)^2\frac{T^3}{v_F^2}
	\end{split}
	\end{equation}\\

	\bibliography{SMRef}{}